\documentclass[12pt]{article}
\title{Axial anomaly and the precise value of the $\pi^0 \to 2 \gamma$ decay width}
\author{B.L.Ioffe and A.G.Oganesian\\ Institute of Theoretical and
Experimental Physics,\\ B.Cheremushkinskaya 25, 117218, Moscow,
Russia}
\date{}
\begin{document}
\maketitle

\newcommand{\be}{\begin{equation}}
\newcommand{\ee}{\end{equation}}
\newcommand{\bea}{\begin{eqnarray}}
\newcommand{\eea}{\end{eqnarray}}
\newcommand{\ve}{\varepsilon}

\def\la{\mathrel{\mathpalette\fun <}}
\def\ga{\mathrel{\mathpalette\fun >}}
\def\fun#1#2{\lower3.6pt\vbox{\baselineskip0pt\lineskip.9pt
\ialign{$\mathsurround=0pt#1\hfil##\hfil$\crcr#2\crcr\sim\crcr}}}

\vspace{1cm}

\begin{abstract}

The anomaly in the vacuum expectation value of the product of
axial  and two vector currents (AVV) in QCD is investigated. The
goal is to determine from its value the $\pi^0 \to 2 \gamma$ decay
width with high precision. The sum rule for AVV formfactor is
studied. The difference $f_{\pi^0} - f_{\pi^+}$ caused by strong
interaction is calculated and appears to be small. The $\pi^0 -
\eta$ mixing is accounted. The $\pi^0 \to 2 \gamma$ decay width
determined theoretically from the axial anomaly is found to be
$\Gamma(\pi^0 \to 2 \gamma) = 7.93 eV$ with an error $\sim 1.5\%$.
The measurement of the $\pi^0 \to 2 \gamma$ decay width at the
same level of accuracy would allow one to achieve a high precision
test of QCD.

\end{abstract}

\hspace{3mm} PACS numbers: 11.15.-q, 11.30.Qc, 12.38.Aw

\newpage

The statement  that in massless QED the axial anomaly is
contributed by massless pseudoscalar state had been first done by
Dolgov and Zakharov in 1971 \cite{Dolgov}. Later this problem was
investigated in many papers. (See, e.g. the book \cite{Peskin} and
the reviews  \cite{Shifman}, \cite{Ioffe}). Now it is known, that
the anomaly for transition of axial isovector current into two
massless vector currents, i.e., into two photons, is almost
completely exhausted by the contribution of $\pi^0 \to 2 \gamma$
decay. However, the accuracy of the calculation of the $\pi^0 \to
2 \gamma$ decay contribution is about 5-7$\%$ \cite{Ioffe} and of
the same order is the experimental error in the determination of
the $\pi^0 \to 2 \gamma$ decay width $\Gamma(\pi^0 \to 2 \gamma)$
\cite{PDG}. The PriMex  experiment at JLab \cite{Kubantsev}, where
it is planned to reduce the error in the experimental value of
$\Gamma(\pi^0 \to 2 \gamma)$ down to 1-2$\%$, is under way. Also,
it is desirable to have theoretical prediction at the same level
of accuracy and therefore, to have a possibility of high precision
test of the anomaly --a significant ingredient of the modern field
theory.

The calculation of hadronic contribution to the AVV anomaly at the
desired level of accuracy was done by Moussalam \cite{Moussallam}
and by Goity, Bernstein and Holstein \cite{Goity} in the framework
of the Chiral Effective Theory (CET) (in the second order) and of
$1/N_c$ expansion. In these calculations the result is expressed
through the parameters of the CET Lagrangian in the second order,
which were determined from the set of the data. In the present
paper we perform such calculation in QCD using the dispersion
relation representation for the AVV formfactor, QCD sum rules to
determine $f_{\pi^0} - f_{\pi^+}$ difference and accounting for
$\pi^0 - \eta$ mixing. The only parameter, which enters the
result, is the value $\Gamma(\eta \to 2 \gamma)$

The notation \be T_{\mu \alpha \beta} (p, p') = \langle p,
\epsilon_{\alpha}; p', \epsilon'_{\beta}  \vert j^{(3)}_{\mu 5}
\vert 0 \rangle \label{1} \ee is used for the matrix element of
the transition of the isovector axial current

\be j^{(3)}_{\mu 5} = (\bar{u} \gamma_{\mu} \gamma_5 u - \bar{d}
\gamma_{\mu} \gamma_5 d)/ \sqrt{2} \label{2} \ee into two photons
with momenta $p$, $p'$ an polarizations $\epsilon_{\alpha}$ ,
$\epsilon'_{\beta}$. (Here $u$ and $d$ are the fields of $u$ and
$d$ quarks.)   The general form of $T_{\mu\alpha \beta}(p, p')$ is
$(q = p+p')$ \cite{Horejsi},\cite{Bass},\cite{Ioffe}:
$$ T_{\mu \alpha \beta}(p, p') = F_1(q^2) q_{\mu}
\epsilon_{\alpha \beta \rho \sigma} p_{\rho} p'_{\sigma} + $$ \be
+ \frac{1}{2} F_2 (q^2) (p_{\alpha} \epsilon_{\mu \beta \rho
\sigma} - p'_{\beta} \epsilon_{\mu \alpha \rho \sigma}) p_{\rho}
p'_{\sigma} \label{3} \ee
 The functions $F_1(q^2)$ and $F_2(q^2)$ can be represented by
 nonsubtracted dispersion relations and in QCD the anomaly
 condition can be written as the sum rule \cite{Horejsi}, \cite{Ioffe}:
\be \int\limits^{\infty}_{(m_u+m_d)^2}~ Im~F_1(q^2) dq^2 =
\sqrt{2} \alpha (e^2_u - e^2_d) N_c, \label{4} \ee where $e_u$ and
$_d$ are $u$-and $d$-quarks electric charges, $e_u = 2/3$, $e_d =
-1/3$, $N_c$ is the number of colours, $N_c=3$. Note, that at
large $q^2$ the function $Im F_1(q^2) \sim (1/q^4) ln q^2$ and the
integral in (\ref{4}) is well converging. Emphasize, that in QCD
there are no perturbative corrections to Eq.(\ref{4}) \cite{Adler}
and it is expected, that nonperturbative corrections are absent
also.

Let us saturate the left-hand side (l.h.s) of (\ref{4}) by the
$\pi^0$ contribution. Use the relation
\be
\langle 0 \vert j^{(3)}_{\mu 5} \vert \pi^0 \rangle = i f_{\pi^0}
q_{\mu} \label{5} \ee The general form of the $\pi^0$-contribution
to $T_{\mu \alpha \beta}(p, p')$ is
\be T_{\mu \alpha \beta} (p, p') = - f_{\pi^0}
\frac{1}{q^2-m^2_{\pi}} A_{\pi} q_{\mu} \epsilon_{\alpha \beta
\lambda \sigma}p_{\lambda} p'_{\sigma} \label{6} \ee where
$A_{\pi}$ is a constant. In the approximation, when only pion
contribution is accounted in the l.h.s. of (\ref{4}) the constant
$A_{\pi}$ is found by substituting (\ref{6}) into (\ref{4}). The
result is (numerical values of $e_u$, $e_d$, $N_c$ were used):
\be
A^{(1)}_{\pi} = \frac{\alpha}{\pi}~ \frac{1}{f_{\pi^0}}. \label{7}
\ee The $\pi^0 \to 2 \gamma$ decay width is easily calculated from
(\ref{6}), (\ref{7}),
\be \Gamma^{(0)} (\pi^0 \to 2 \gamma) = \frac{\alpha^2}{32\pi^3}~
\frac{m^3_{\pi^0}}{f^2_{\pi^0}} \label{8}\ee Index $(0)$ at
$\Gamma$ means that the saturation of the sum rule (\ref{4}) by
the $\pi^0$ state was exploited. At $f_{\pi^0} = f_{\pi^+} = 130.7
\pm 0.4$ MeV and $m_{\pi^0} = 135.0$ MeV \cite{PDG} we get from
(\ref{8}):
\be
\Gamma^{(0)} (\pi^0 \to 2 \gamma) = 7.73 eV \label{9} \ee

Turn now to the calculation of correction to zero approximation.
The first correction is due to the fact, that generally
$f_{\pi^0}$ is not equal to $f_{\pi^+}$.  There are two sources of
the $f_{\pi^0}-f_{\pi^+}$ difference: the electromagnetic
interaction and violation of isospin in strong interaction.\\
 $\pi^0$ has no electromagnetic interaction and the
 electromagnetic interaction of $\pi^+$ was already accounted,
 when the value of $f_{\pi^+}$ was found from the $\pi^+ \to \mu^+
 \nu$ decay data \cite{PDG}. So, the electromagnetic interaction
 does not change the value of $f_{\pi^0}$ in comparison with the
 presented above value of $f_\pi^+$. (The discussion of magnitude of
  $f_{\pi^0}-f_{\pi^+}$ difference, caused by electromagnetic interaction
  was done in Ref.\cite{Goity}. In order to find $f_{\pi^0} -
 f_{\pi^+}$ caused by strong interaction consider the polarization
 operators $\Pi^{(3)}_{\mu \nu}$ and $\Pi^{(+)}_{\mu \nu}$ of
 axial currents $j^{(3)}_{\mu 5}$ (\ref{2}) and $j^{(+)}_{\mu 5}
 = \bar{u} \gamma_{\mu} \gamma_5 d$. (The presented below method
 is exposed in \cite{BIoffe}, \cite{Geshkenbein} and reviewed in \cite{BLIoffe}).
 The general form of these polarization operators is
 \be
 \Pi^{(i)}_{\mu \nu} (q) = \Pi^{(i)}_T(q^2)(q_{\mu} q_{\nu} -
 \delta_{\mu \nu} q^2) + q_{\mu} q_{\nu} \Pi^{(i)}_L (q^2), ~~ i =
 3, +
 \label{10}
 \ee
 The pion contribution to $\Pi_{\mu \nu}(q)$ is given by \cite{BIoffe}:
 \be
 \Pi_{\mu \nu} (q)_{\pi} = -\frac{f^2_{\pi}}{q^2} (q_{\mu} q_{\nu}
 - \delta_{\mu \nu} q^2) - \frac{m^2_{\pi}}{q^2} q_{\mu} q_{\nu} ~
 \frac{f^2_{\pi}}{q^2 - m^2_{\pi}} .
 \label{11}
\ee Consider the longitudinal polarization operator
$\Pi^{(i)}_L(q^2)$ to which only pseudoscalar mesons are
contributing. In order to separate the interesting for us second
term in (\ref{11}) let us multiply (\ref{11}) by
$q_{\mu}q_{\nu}/q^2$ and consider the difference
\be q^2\biggl ( \Pi^{(3)}_L(q^2) - \Pi^{(+)}_L(q^2)\biggr ) = -
\frac{m^2_{\pi^0}f^2_{\pi^0}}{q^2 - m^2_{\pi^0}} +
\frac{m^2_{\pi^+}f^2_{\pi^+}}{q^2 - m^2_{\pi^+}} \label{12} \ee As
was demonstrated by Weinberg \cite{Weinberg}, in the first order
in $m_u - m_d$ the $\pi^+$ and $\pi^0$ masses are equal and the
experimentally observed $\pi^+$ and $\pi^0$ mass difference arises
from electromagnetic interaction. In the second order in $u$- and
$d$-quark masses $m^2_{\pi^0} - m^2_{\pi^+}$ is proportional to
$(m_u-m_d)^2$ or $(m_u - m_d)~ [\langle 0 \vert \bar{u} u \vert 0
\rangle - \langle 0 \vert \bar{d} d \vert 0 \rangle ]$ and, as can
be shown, is very small. So, in (\ref{12}) we can put $m^2_{\pi^+}
= m^2_{\pi^0}$ and have
\be \Delta(q^2) \equiv q^2\biggl (\Pi^{(3)}_L (q^2) - \Pi^{(+)}_L
(q^2)\biggr ) = - \frac{m^2_{\pi}}{q^2 - m^2_{\pi}} (f^2_{\pi^0} -
f^2_{\pi^+}). \label{13} \ee

Exploit the standard QCD sum rule technique and represent
$\Delta(q^2)$ as an operator product expansion (OPE)
\be \Delta(q^2) = R_2(q^2) + R_4(q^2) + R_6(q^2), \label{14} \ee
where $R_2$ corresponds to the contribution of bare loop diagram,
$R_4$ -- to the operator of dimension 4 and $R_6$ -- to the
operator of dimension 6. Only the terms, proportional to $m_u -
m_d$ and $\langle 0 \vert \bar{u} u \vert 0 \rangle - \langle 0
\vert \bar{d} d \vert 0 \rangle$ remain in (\ref{14}). Note, that
$\Delta(q^2)$ is even under interchange $u \leftrightarrow d$. For
this reason no linear in $m_u - m_d$ or $\langle 0 \vert \bar{u} u
\vert 0 \rangle - \langle 0 \vert \bar{d} d \vert 0 \rangle$ terms
can survive in the right-hand side of (\ref{14}). Calculating the
terms of OPE we get after Borel transformation:
$$ \frac{m^2_{\pi}}{M^2}e^{-m^2_{\pi}/M^2}(f^2_{\pi^0} - f^2_{\pi^+}) \approx
\frac{m^2_{\pi}}{M^2} (f^2_{\pi^0} - f^2_{\pi^+}) = \frac{3}{8
\pi^2} (m_u - m_d)^2 - $$ \be - \frac{m_d - m_u}{4 \pi^2 M^2} (a_u
- a_d), \label{15} \ee where
\be
a_q = -(2 \pi)^2 \langle 0 \vert \bar{q} q \vert 0 \rangle, ~~ q =
u,d \label{16} \ee and $M^2$  is the Borel parameter. The
contribution of the $d=6$ operator vanishes. For the difference
$f_{\pi^0} - f_{\pi^+}$ we have
\be \frac{\Delta f_{\pi}}{f_{\pi}} \equiv \frac{f_{\pi^0} -
f_{\pi^+}}{f_\pi} = \frac{3}{16 \pi^2}~ \frac{M^2}{m^2_{\pi}
f^2_{\pi}} \Biggl [(m_u - m_d)^2 - \frac{2}{3}~ \frac{m_d -
m_u}{M^2 } \gamma a_q \Biggr ], \label{17} \ee where
\be
\gamma = \frac{a_u - a_d}{a_q} \label{18} \ee

For numerical estimations we put $m_d - m_u = 3.5 MeV \pm 20\%$
\cite{BLIoffe}, $\gamma = 6\cdot 10^{-3}$ \cite{Gasser}  or
$\gamma = 3\cdot 10^{-3}$ \cite{Adami} and $M = 1 GeV$. The
calculations give
\be \frac{\Delta f_{\pi}}{f_{\pi}} \simeq 2 \cdot 10^{-4} ~~~
\cite{Gasser}; ~~~~~~ \frac{\Delta f_{\pi}}{f_{\pi}} \simeq 4
\cdot 10^{-4} ~~ \cite{Adami} \label{19} \ee Therefore, the
difference $f_{\pi^0} - f_{\pi^+}$ is negligible.

The additional source of the $f_{\pi^0} - f_{\pi^+}$ difference
arises because there is the contribution of $\eta$-meson to the
polarization operator $\Pi^{(3)}_L$, besides the $\pi^0$. This
contribution appears from the $\eta-\pi$ mixing and must be
subtracted from the value of $f_{\pi^0}$. Omitting the details of
simple calculations, we present the result
\be \Biggl ( \frac{\Delta f_{\pi}}{f_{\pi}} \Biggr ) = -
\frac{1}{2}~ \frac {m^2_{\eta}}{m^2_{\pi}}~ Sin^2 \theta_{\eta
\pi}~ \exp[-(m^2_{\eta} - m^2_{\pi})/M^2] \label{20} \ee where
$\theta{\eta \pi}$ is the $\eta-\pi$ mixing angle given below
(Eq.(\ref{29})). Numerically
\be \Biggl ( \frac{\Delta f_{\pi}}{f_{\pi}} \Biggr )= -
1.3.10^{-3} \label{21} \ee and is also very small.

Let us dwell now on the main correction to the zero approximation
result for $\Gamma(\pi^0 \to 2 \gamma)$ arising from excited
states contributions to the sum rule (\ref{4}). The next after
$\pi^0$ pseudoscalar meson is $\eta$. It can contribute to
(\ref{4}) because of isospin violation, caused by different masses
of $u$- and $d$- quarks and resulting in $\pi^0 - \eta$ mixing.
The problem of $\pi^0 - \eta$ mixing in QCD was considered in
\cite{IoffeBL}, \cite{Gross}, \cite{ShifmanM}. Following
\cite{IoffeBL}, \cite{ShifmanM} introduce nonorthogonal states
$\vert P_3 \rangle$ and $\vert P_8 \rangle$ and the corresponding
fields $\varphi_3$, $\varphi_8$, coupled to $j^{(3)}_{\mu 5}$ and
$j^{(8)}_{\mu 5}$:
\be \langle 0 \vert j^{(k)}_{\mu 5}\vert P_l \rangle = i
\delta_{kl} f_k q_{\mu}, ~~ k = 3,8. \label{22} \ee
Nonorthogonality of the fields $\varphi_3$, $\varphi_8$
corresponds to the non-diagonal term $\Delta H = m^2_{\eta \pi}
\varphi_3 \varphi_8$ in the effective interaction Hamiltonian. In
the presence of such term the standard PCAC relations are modified
in the following way \cite{IoffeBL}
$$
\partial_{\mu} j^{(3)}_{\mu 5} = f_{\pi}(m^2_{\pi} \varphi_3 +
m^2_{\eta \pi} \varphi_8), $$ \be
\partial_{\mu} j^{(8)}_{\mu 5} = f_{\eta} (m^2_{\eta} \varphi_8 +
m^2_{\eta \pi} \varphi_3). \label{23} \ee The fields $\varphi_3$,
$\varphi_8$ are expressed through the physical fields
$\varphi_{\pi}$, $\varphi_{\eta}$ as
$$ \varphi_3 = Cos \theta_{\eta \pi} \varphi_{\pi} + Sin
\theta_{\eta \pi} \varphi_{\eta} $$ \be \varphi_8 = - Sin
\theta_{\eta \pi} \varphi_{\pi} + Cos \theta_{\eta \pi}
\varphi_{\eta}, \label{24} \ee where the mixing angle is given by
\be \theta_{\eta \pi} \approx \frac{m^2_{\eta \pi}}{m^2_{\eta} -
m^2_{\pi}} \approx \frac{m^2_{\eta \pi}}{m^2_{\eta}}. \label{25}
\ee

In QCD the nondiagonal mass $m^2_{\eta\pi}$ is expressed through
the difference of $u$- and $d$-quark masses
\cite{IoffeBL},\cite{Gross}: \be m^2_{\eta\pi}
=\frac{1}{\sqrt{3}}\frac{m_u-m_d}{m_u+m_d} m^2_{\pi}.\label{26}\ee
Now $Im~F_1(q^2)$ is given by the sum of contributions of $\pi^0$
and $\eta$ mesons. In order to separate the formfactor $F_1(q^2)$
and to kill the contributions of axial mesons in the
representation of $Im~T_{\mu\alpha\beta}(p,p^{\prime})$ in terms
of physical states, multiply Eq.(\ref{3}) by $q_{\mu}/q^2$. Using
Eq.'s (\ref{1}),(\ref{23}),(\ref{24}) and taking the imaginary
part, we get:
$$ Im~ q_{\mu}\frac{1}{q^2} \langle 2\gamma \mid j^{(3)}_{\mu 5}\mid
0 \rangle = -\frac{1}{q^2} Im \langle 2\gamma \mid m^2_{\pi} (Cos
\theta_{\eta\pi} \varphi_{\pi} +Sin \theta_{\eta\pi}
\varphi_{\eta}) +$$ $$ +m^2_{\eta\pi} (-Sin \theta_{\eta\pi}
\varphi_{\pi} + Cos \theta_{\eta\pi} \varphi_{\eta})\mid 0 \rangle
\approx \pi [~\delta (q^2-m^2_{\pi}) A_{\pi} +$$
$$+ \frac{m^2_{\pi}}{m^2_{\eta}} \frac{m^2_{\eta\pi}}{m^2_{\eta}} \delta
(q^2-m^2_{\eta}) A_{\eta} - \frac{(m^2_{\eta\pi})^2}{m^2_{\pi}
m^2_{\eta}} \delta(q^2-m^2_{\pi}) A_{\pi}
+\frac{m^2_{\eta\pi}}{m^2_{\eta}} \delta (q^2
-m^2_{\eta})A_{\eta}~]\approx$$ \be\approx \pi \left\{ \delta
(q^2-m^2_{\pi})A_{\pi} \Biggl [
1-\frac{m^2_{\pi}}{m^2_{\eta}}\frac{1}{3} \Biggl
(\frac{m_u-m_d}{m_u+m_d}\Biggr )^2 \Biggr ] +\frac{1}{\sqrt{3}}
\frac{m^2_{\pi}}{m^2_{\eta}} \Biggl (\frac{m_u-m_d}{m_u+m_d}\Biggr
) \delta(q^2 -m^2_{\eta}) A_{\eta}\right\},\label{27}\ee where
$A_{\eta}$ is the amplitude $\eta\to 2\gamma$ decay. The ratio
$A_{\eta}/A_{\pi}$ is equal \be \frac{A_{\eta}}{A_{\pi}} =\biggl
[\frac{\Gamma(\eta\to 2\gamma)}{\Gamma(\pi^0\to
2\gamma)}\frac{m^3_{\pi}}{m^3_{\eta}}\biggr ]^{1/2}.\label{28}\ee
Numerically, at $\Gamma(\eta\to 2\gamma)=510 eV$ \cite{PDG} and
$\Gamma(\pi^0 \to 2\gamma)=7.73 eV$, $A_{\eta}/A_{\pi}=1.0$.
According to (\ref{25}),(\ref{26}) the mixing angle
$\theta_{\eta\pi}$ is given by (the numerical data were taken from
Ref.\cite{BLIoffe}): \be \theta_{\eta\pi} =\frac{1}{\sqrt{3}}
\frac{m_u-m_d}{m_u+m_d} \frac{m^2_{\pi}}{m^2_{\eta}} =- 0.0150 \pm
0.020\label{29}\ee and the total correction to the amplitude of
$\pi^0\to 2\gamma$ decay arising from $\eta-\pi$ mixing is \be
\frac{(F_1)_{\eta}}{(F_1)_{\pi}} = -(1.2 \pm 0.25)\%.\label{30}\ee

Let us estimate the contributions to the sum rule (\ref{4}) from
higher pseudoscalar mesons. The $\eta^{\prime}$ meson contributes
through $\eta\eta^{\prime}$ mixing. Its contribution is suppressed
by the ratio $(m_{\eta}/m_{\eta^{\prime}})^2$ in comparison with
$\eta$ contribution and we have \be
\frac{(F_1)_{\eta^{\prime}}}{(F_1)_{\eta}} \sim
\sqrt{\frac{\Gamma(\eta^{\prime}\to 2\gamma)}{\Gamma(\eta\to
2\gamma)}}\Biggl (\frac{m_{\eta}}{m_{\eta^{\prime}}}\Biggr )^{3/2}
\Biggl (\frac{m_{\eta}}{m_{\eta^{\prime}}}\Biggr )^2
Sin\vartheta_{\eta\eta^{\prime}}, \label{31}\ee where
$\vartheta_{\eta\eta^{\prime}}$ is the $\eta\eta^{\prime}$ mixing
angle. In accord with \cite{Feldmann} we can put
$\vartheta_{\eta\eta^{\prime}}\approx -20^0$. Using
$\Gamma(\eta^{\prime}\to 2\gamma) =4.29 MeV$ \cite{PDG} we have
the estimation \be (F_1)_{\eta^{\prime}}/(F_1)_{\eta} \sim
0.15\label{32}\ee Finally, estimate the contribution of the
resonance $\pi(1300)$. The contribution of this resonance to
$T_{\mu\alpha\beta}(p,p^{\prime})$ (\ref{1}) (as well as any other
excited states) should be proportional to $m_u +m_d$ or, what is
equivalent, to $m^2_{\pi}$, otherwise the axial current would not
conserve in the limit $m_u+m_d\to 0$. Therefore, for dimensional
grounds, \be \frac{(F_1)_{\pi(1300)}}{(F_1)_{\pi}} \sim
\frac{m^2_{\pi}}{m^2_{\pi(1300)}} \cdot < 1 \%. \label{33}\ee In
fact, probably, (\ref{33}) overestimates the $\pi(1300)$
contribution, since, as was mentioned above, $F_1(q^2)$ decreases
as $1/q^4$ at large $q^2$.  So, in what follows we take
$(F_1)_{\pi(1300)}/(F_1)_{\pi} \sim 0.5\%$. Summing all
uncertainties in squares we have the total uncertainty in the
amplitude about 0.7\%. The final result for $\Gamma(\pi^0\to
2\gamma)$ from the axial anomaly is (the correction, given by
Eq.(\ref{21}) is accounted): \be \Gamma(\pi^0\to 2\gamma) = 7.93
eV \pm 1.5\%.\label{34}\ee In the limit of errors this agrees with
those found in \cite{Moussallam} and \cite{Goity}. The
experimental test of this relation, which is planned by PriMex
Collaboration would be very important: it would be a high
precision test of QCD and even more general -- the phenomenon of
the anomaly. Not too many such high precision tests of QCD are
known till now.

We  are thankful to A.G.Dolgolenko, who informed us about the
experiment PriMex, what stimulated this work. This work was
supported in part by US CRDF Cooperative Grant Program, Project
RUP2-2621-MO-04,  RFBR grant 06-02-16905a and the funds from EC to
the project ``Study of Strongly Interacting Matter'' under
contract No. R113-CT-2004-506078.

\end{document}